\documentclass[aps,12pt]{revtex4}
\usepackage[dvips]{epsfig}

\parskip=\medskipamount

\def\be{\begin{equation}}
\def\ee{\end{equation}}
\def\disp{\displaystyle}

\def\R{{\sf I\kern-.15em R}}
\def\T{{\sf T\kern-.45em T}}
\def\C{\kern.1em{\raise.47ex\hbox{$\scriptscriptstyle |$}}
             \kern-.40em{\sf C}}
\def\Z{{\sf Z\kern-.45em Z}}

\begin{document}

\title
{\Large Topological correlations in trivial knots: new arguments in support
of the crumpled polymer globule}

\author
{S.K. Nechaev$^{1,2}$, O.A. Vasilyev$^{1}$}

\affiliation
{\it $^1$Landau Institute for Theoretical Physics,
117334 Moscow, Russia \\
$^2$UMR 8626, CNRS--Universit\'e Paris XI, LPTMS, Universit\'e Paris
Sud, 91405 Orsay Cedex, France}

\date{\today}

\begin{abstract}

We prove the fractal crumpled structure of collapsed unknotted polymer
ring. In this state the polymer chain forms a system of densely packed
folds, mutually separated in all scales. The proof is based on the
numerical and analytical investigation of topological correlations in
randomly generated dense knots on strips $L_{v} \times L_{h}$ of widths
$L_{v}=3,5$. We have analyzed the conditional probability of the fact that
a part of an unknotted chain is also almost unknotted. The complexity of
dense knots and quasi--knots is characterized by the power $n$ of the
Jones--Kauffman polynomial invariant. It is shown, that for long strips
$L_{h} \gg L_{v}$ the knot complexity $n$ is proportional to the length of
the strip $L_{h}$. At the same time, the typical complexity of the
quasi--knot which is a part of trivial knot behaves as $n\sim \sqrt{L_{h}}$
and hence is significantly smaller. Obtained results show that topological
state of any part of the trivial knot in a collapsed phase is almost
trivial.

\end{abstract}

\maketitle

\section{Introduction}
\label{sect:1}

It is well known, that very often new interesting problems appear
at the edges of traditional fields of science. Statistical topology,
emerged from statistical physics, theory of integrable systems and
algebraic topology can be regarded as an example of such new joint area.
The scope of problems dealing with construction of knot and link
invariants as well as study of entropy of random knots belongs to
this field.

In this paper the modern methods of algebraic topology are used to argue
for the nontrivial fractal structure of unknotted polymer ring in a compact
state. Namely, we show that the condition for the whole knot to be trivial
implies that each part of such knot in the compact state is almost
unknotted.

Let us remind that the non-phantomness of polymer chains causes two types
of interactions: a) volume interactions, vanishing for infinitely thin
chains, and b) topological interactions, remaining even for chains of zero
thickness. For sufficiently high temperatures the polymer macromolecule is
strongly fluctuating system without reliable thermodynamic state. However
for temperatures below some critical value $\theta$ the macromolecule forms
a dense drop--like weakly fluctuating globular structure \cite{lif,lgkh}.
In classical works \cite{lif,lgkh} devoted to investigation of
coil--to--globule phase transition without topological constraints it has
been shown, that for $T<\theta$ the globular state can be described in
virial approximation, i.e.  using only two-- and three--body interaction
constants: $B=b\frac{T-\theta} {\theta}<0$ and $C={\rm const}>0$ (see also
\cite{grosbk}). The approach developed in \cite{lif,lgkh} is considered as
the basis of the modern statistical theory of collapsed state of polymer
systems.

\begin{figure}[ht]
\begin{center}
\epsfig{file=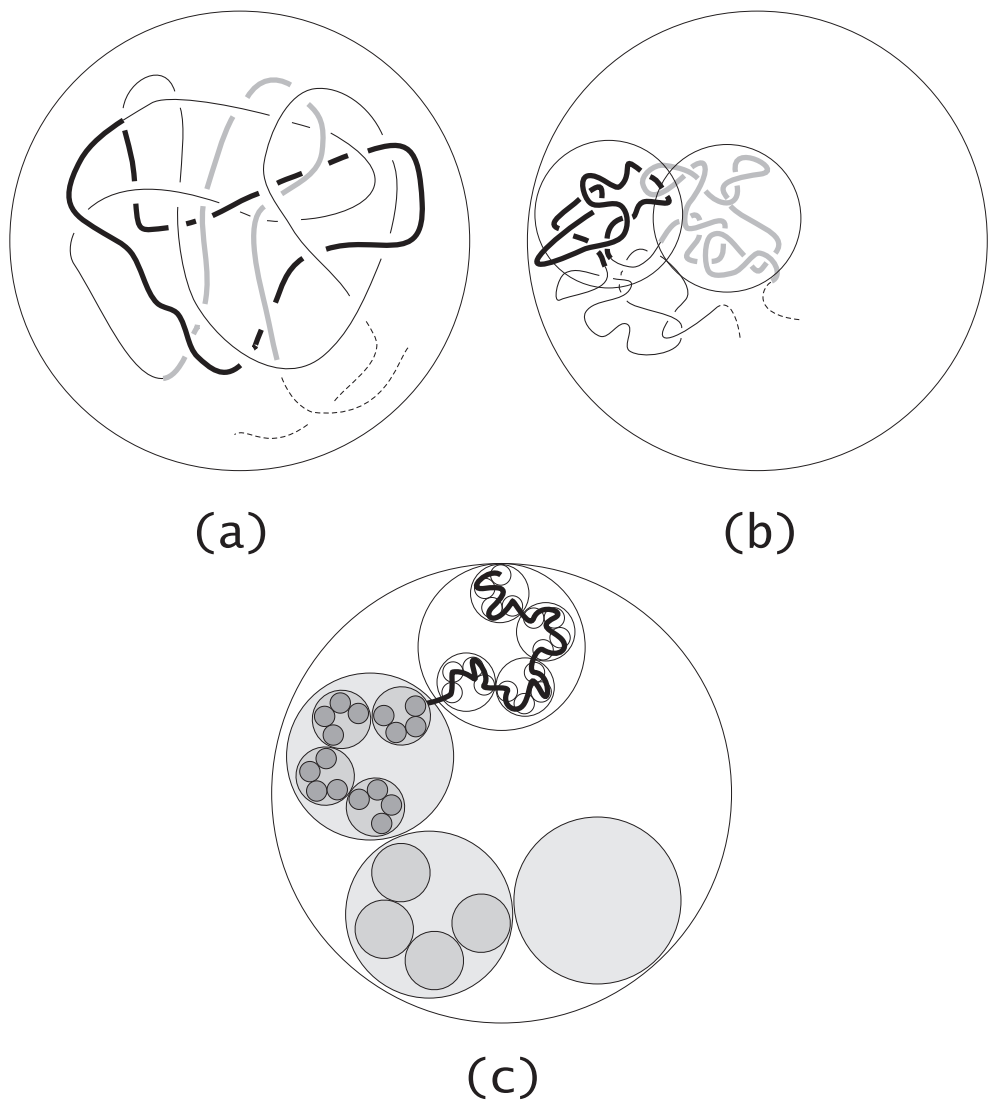,width=13cm}
\end{center}
\caption{a) The globular (collapsed) state of a polymer chain with an arbitrary
loop penetrating to the size of a whole globule; b) The crumpled globular state
with a test loop segregated "in itself". This structure resembles the Peano
curve schematically shown in c).}
\label{fig:0}
\end{figure}

In the globular phase of unknotted macromolecule the topological constraints
play a role of an additional repulsion. One might expect that the parts of
the unknotted chain deeply penetrate each other by loops as it is shown in
Fig.\ref{fig:0}a. However this is not true and in the paper \cite{gns} it
has been shown that the absence of knots on densely packed polymer ring
causes very unusual fractal properties of the chain trajectory strongly
affecting all the thermodynamic properties of the macromolecule in the
globular phase. The corresponding structure of a collapsed unknotted
polymer ring was named "crumpled" fractal globule. The chain trajectory in
the crumpled globule densely fills the volume such that all parts of the
chain become segregated from each other in a broad region of scales---see
Fig.\ref{fig:0}. Hence, the line path in a crumpled globule reminds the
well known Peano curve \cite{man} schematically shown in Fig.\ref{fig:0}c.

Experimental examination of a fractal structure of unknotted polymer ring
is a very difficult problem. Some measurements can be interpreted  as the
indirect verification of the crumpled globule structure: a two--stage
dynamics of a collapse of a linear macromolecule after abrupt change of a
solvent quality \cite{chu} and the effect of compatibility enhancement in a
melt of linear and ring macromolecules \cite{mcknight}. However now there
was no direct observation of crumpled globule structure in equilibrium
globular polymer chains.

We show below that the investigation of distribution of random knots over the
topological classes and analysis of topological correlations in trivial knots
helps one to understand the structure of the phase space of unknotted
polymer in a globule phase and validate the crumpled globule conjecture
"from the first principle".

\section{The model of dense knots and a concept of "quasi--knots"}
\label{sect:2}

Below we express the arguments in support of the crumpled globule  on the basis
of direct determination of the topological state of a {\it part} of a polymer
chain under the condition that the chain as a whole forms an unknotted loop. In
our work we use some modern methods of construction of topological knot
invariants \cite{jones,kauffman,wu} applied for solution of specific
statistical problems \cite{grne_alg,vasne}.

First of all one has to define the topological state of a part of a ring
polymer chain. Certainly, the mathematically rigorous definition of the
topological state exists for closed or infinite paths. Nevertheless, the
everyday's experience tells us that open but sufficiently long rope can be
knotted. Hence, it is desirable to introduce the concept of a {\it quasi--knot}
available for topological description of open paths.

For the first time that idea in the polymer context had been formulated by
I.M. Lifshits and A.Yu. Grosberg \cite{ligr}. They argued that the  topological
state of a linear polymer chain in a globular state is defined  much better
than topological state of a coil. Actually, the distance between the ends of
the chain in a globule is of order $R \sim a N^{1/3}$, where $a$ is a size of a
monomer and $N$ is a number of monomers in a chain. Taking into  account that
$R$ is sufficiently smaller than the contour length $N$ and that the density
fluctuations are negligible, we may define the topological state of a path in a
globule as a topological state of composition of a chain itself and a segment
connecting its ends. This composite structure can be regarded as a quasi--knot
for an open chain in a  globular state. Later on we shall repeatedly use this
definition.

\begin{figure}[ht]
\begin{center}
\epsfig{file=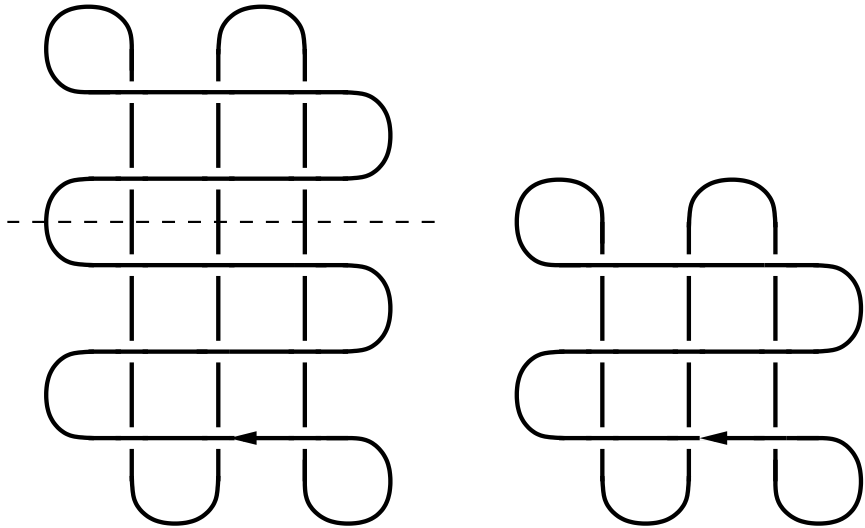,width=7cm,angle=90}
\end{center}
\caption{The trivial knot $3 \times 5$ and the knot $3 \times 3$ obtained by
cutting a knot $3 \times 5$.}
\label{fig:1a}
\end{figure}

Let us describe the model under consideration. The crumpled globule is modeled
by "dense" knots. We call the knot "dense" if its projection onto  the plane
fully fills the rectangle lattice $M$ of size $L_{v} \times  L_{h}$
--- see Fig.\ref{fig:1a}.

The lattice $M$ is filled densely by a single thread, which crosses itself in
each vertex of the lattice in two different ways: "above" or "below". The
topology of a knot is defined by "above" and "below"  passages with prescribed
boundary conditions. The "woven carpet" shown in Fig.\ref{fig:1a} corresponds
to a trivial knot. Let use enumerate the  vertices of the lattice by the index
$k$ and attribute to each vertex the  variable $\epsilon_{k}$ according to the
rule:

\unitlength=1.00mm
\special{em:linewidth 0.4pt}
\linethickness{0.4pt}
\begin{picture}(110.00,20.00)
\put(10.00,10.00){\line(1,0){8.00}}
\put(22.00,10.00){\vector(1,0){8.00}}
\put(20.00,0.00){\vector(0,1){20.00}}
\put(80.00,10.00){\vector(1,0){20.00}}
\put(90.00,0.00){\line(0,1){8.00}}
\put(90.00,12.00){\vector(0,1){8.00}}
\put(40.00,10.00){\makebox(0,0)[cc]{$\epsilon=-1$}}
\put(110.00,10.00){\makebox(0,0)[cc]{$\epsilon=+1$}}
\end{picture}

Our task can be rephrased now as follows. We have an ensemble of randomly
generated crossings $\{\epsilon_k\}$
$$
\left\{\begin{array}{ll}
\epsilon_k=-1 & \mbox{with probability $p$} \\
\epsilon_k=+1 & \mbox{with probability $1-p$}
\end{array} \right.
$$
independent in all vertices under the condition that the total set of
crossings $\{\epsilon_k\}$ defines a trivial knot. We call this initial
knot the "parent" one. If $\epsilon_{k}=+1$ for all $k=[1,N]$ then the knot
is trivial. Changing the sign for some $\epsilon_{k}$ from "$+$" to "$-$"
we can get a nontrivial knot.  Later we shall call such change of a sign
"impurity", so $p$ is concentration of impurities. Let us cut a part of the
parent trivial knot closing the open ends of the threads as it is shown in
Fig.\ref{fig:1a}. This  way we get the well--defined "daughter"
quasi--knot. Below we study the typical topological state  of such daughter
quasi--knots under the condition that the parent knot is trivial.

The model under consideration is oversimplified because it does not take into
account the spatial fluctuations of a path of a polymer chain in a knot.
However we believe that our model well describes the globular (condensed)
structure of a polymer ring where the chain fluctuations are essentially
suppressed and the polymer has reliable thermodynamic structure with almost
constant density \cite{ligr}.

\section{Topological invariants and knot complexity}
\label{sect:3}

In our work we characterize the topological state of the knot by its
Jones--Kauffman polynomial invariant \cite{jones,kauffman}. Recall that the
Kauffman invariant of variable $A$ is a Jones polynomial of variable $x=A^{4}$
\cite{kauffman}.

\begin{figure}[ht]
\begin{center}
\epsfig{file=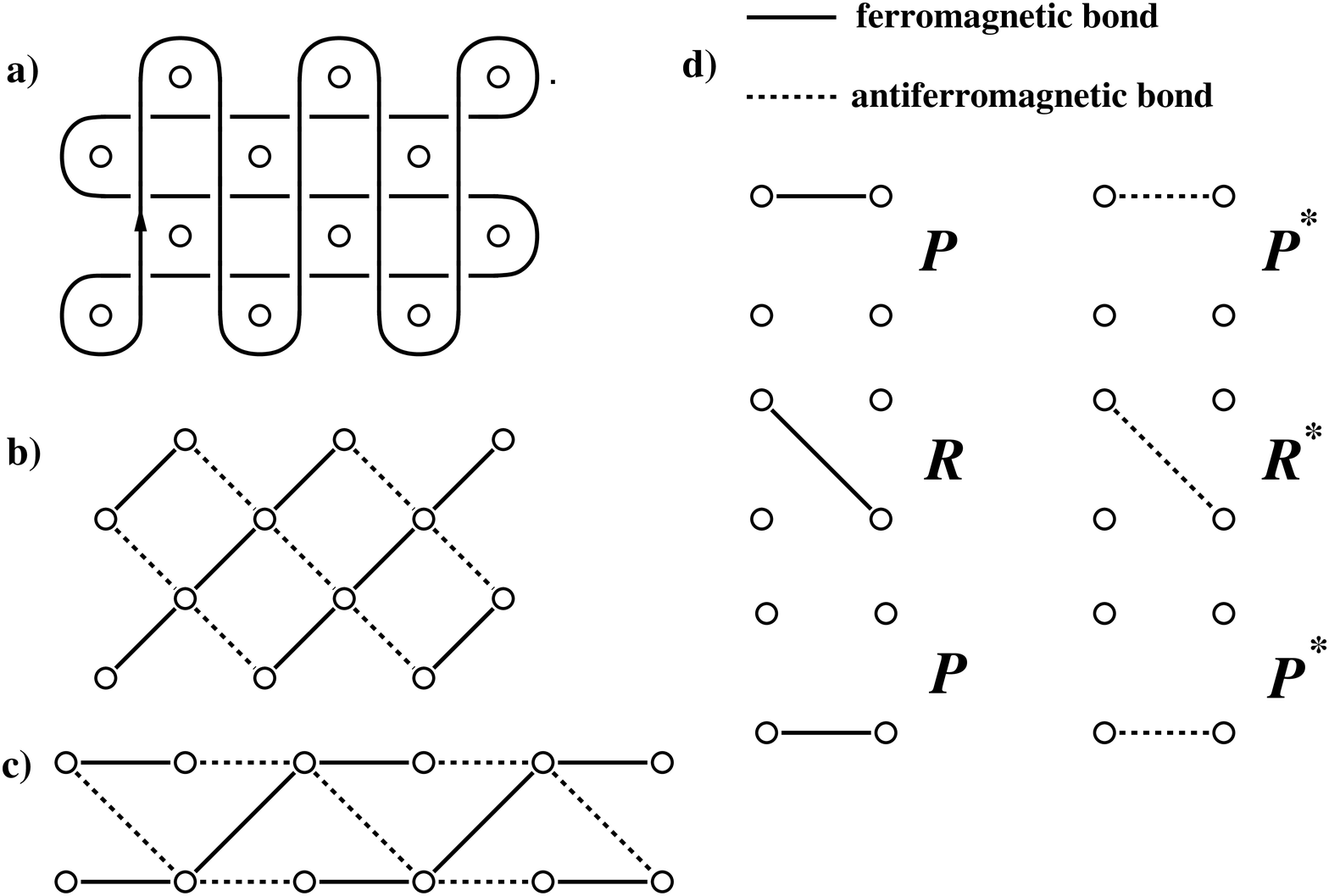,width=15cm}
\end{center}
\caption{a) The trivial knot $3\times 5$; b) The lattice of Potts spins,
corresponding to the trivial knot $3 \times 5$; c) The lattice of Potts
spins transformed to $2 \times 6$---strip corresponding to the trivial knot
$3 \times 5$; d) The transfer matrix, corresponding to ferro-- and
antiferro--magnetic bonds on the Potts lattice.}
\label{fig:1b}
\end{figure}

In brief the procedure of construction of the algebraic knot invariant is
as follows. Let us construct the "Potts lattice" $\Lambda$ corresponding to
the "knot lattice" $M$---see Fig.\ref{fig:1b}b.
\bigskip

In Fig.\ref{fig:1b}a positions of Potts spins are depicted by circles.
Introduce the set of  variables $b_{i,j}$ on the bonds of the Potts lattice as
follows:
\be \label{eqbb}
b_{i,j}= \left\{
\begin{array}{cl}
-\epsilon_{k}, & \mbox{for right--down bond ($i,j$)} \\
\epsilon_{k}, & \mbox{for right--up bond ($i,j$)}
\end{array} \right.
\ee
We can treat $b_{i,j}=1$ and $b_{i,j}=-1$ as "ferromagnetic" and
"antiferromagnetic" bonds, because changing a sign $b_{i,j}$ means changing
a sign of the coupling constant $J_{i,j}$. Let us note, that even for the
trivial knot we have antiferromagnetic bonds on the Potts lattice --- see
Fig.\ref{fig:1b}a and Fig.\ref{fig:1b}b,c. We should stress that in our
terms the "impurity" means not an antiferromagnetic bond, but just the
change of the sign of a bond with respect to the trivial knot. In
Fig.\ref{fig:1b}b and Fig.\ref{fig:1b}c the ferromagnetic $b_{i,j}=1$ and
antiferromagnetic $b_{i,j}=-1$ bonds are shown by solid and dashed lines
correspondingly.

In \cite{grne_alg,vasne} it has been explicitly shown that the Jones knot
invariant $f_K(x)$ of variable $x$ of ambiently isotopic knots on the
lattice $M$ can be expressed as a partition function of the Potts model on
the lattice $\Lambda$:
\be \label{eqkp2}
f_K(x,\{b_{i,j}\})=K \left(x,\{b_{i,j}\}\right)
Z\Big(q(x),\{J_{i,j}(b_{i,j},x) \}\Big)
\ee
where
\be \label{eqh}
K\left(x,\{b_{i,j}\}\right) =
\left( -\sqrt{x} -\frac{1}{\sqrt{x}} \right)^{-(N_{p}+1)}
\left(-\sqrt{x} \right)^{ \sum\limits_{\{ i,j\} } b_{i,j} }
\ee
is a trivial prefactor, independent on Potts spins and $N_{p}=\frac{1}{2}
(L_{v}+1)(L_{h}+1)$ is a total number of Potts spins.

The Potts partition function $Z\Big(q(x),\{J_{i,j}(b_{i,j},x)\}\Big)$
defined in (\ref{eqkp2}) reads:
\be \label{eqpb}
Z\Big(q(x),\{J_{i,j}(b_{i,j},x)\}\Big)=\sum \limits_{\{\sigma\}}
\exp \left(\sum \limits_{\{i,j\}}\frac{J_{i,j}(b_{i,j},x)}{T}
\delta(\sigma_{i},\sigma_{j})\right)
\ee
where the coupling constant $J_{i,j}$ and the number of spin states $q$
are:
\be \label{eqjq}
\frac{J_{i,j}}{T}=\ln(-x^{b_{i,j}}), \quad q=x+2+x^{-1}
\ee
The specific property of the partition function (\ref{eqpb}) consists in the
relation between the temperature $T$ and the number of the Potts spins states,
$q$, so the variables $T$ and $q$ cannot be considered as independent
quantities. Later on we shall be interested in the largest  power of the Jones
polynomial invariant, hence we forget about specific physical sense of the
parameters $x$, $T$, and $J_{i,j}$.

In the paper \cite{vasne} we have investigated semi--numerically the
probability $P\{f_K\}$ that among $2^N$ possible realizations of the disorder
$\{\epsilon_{k}\},\;k=1,\dots,N$ one can find a knot in a specific topological
state defined by the Jones--Kauffman invariant $f_{K}$.
$$
P_N\{f_K\}=\frac{1}{2^N} \sum \limits_{\{\epsilon_{k}\}} \Delta
\Big(f_K(x,\{\epsilon_{1}, \dots,\epsilon_{N}\})-f_K\Big)
$$

The topological disorder described by random variables $\epsilon_k= \pm 1$ can
be considered as a quenched random disorder in an  appropriate Potts
model. Each topological type of a knot is characterized  by polynomial
invariant $f_K(x)$. However it is sufficient and more convenient for our
particular goals  to discriminate knots by more rough characteristics -- the
power $n$ of the polynome $f_K(x)$:
\be \label{eq:exp}
n=\lim_{|x|\to\infty}\frac{\ln f_K(x)}{\ln x}
\ee
For trivial knots $n=0$; as the knot becomes more complex, $n$ grows up
to its maximal value $n \sim N$.

Certainly, the lack of our description consists in the fact that $n$ is rather
rough characteristic and one cannot distinguish "topologically similar" knots.
However the discrimination of the topological state of knots (or links) by the
power $n$ of a corresponding polynome enables us to introduce a "metric" in the
phase space of topological states and, hence, to compare the knots.

\section{Topological correlations of (quasi)knots on the strip}
\label{sect:4}

In this section by numerical and analytical methods we investigate the typical
topological state of a part of a "parent" knot under the condition that the
parent knot is trivial. In fact, we slightly modify the very problem as soon as
we characterize the topological state of the knot by the power $n$ of its
algebraic polynome. Namely, we study topological correlations not only in
trivial knots, but in all knots characterized by zero's power of the knot
polynome as well. There are some non-trivial knots among a variety of knots
with $n=0$, but the complexity of such non-trivial knots is not high. We could
say that the knots with $n=0$ are "about trivial".

In our numerical investigation we restrict ourselves to topological
correlations in knots represented by strips of sizes $L_v\times L_h$, where
$L_v$ is the width ($L_v=3,\,5$) and $L_{h}$ is the length of the strip. We
consider essentially asymmetric distribution of crossings, namely with
$p=0.05,0.1$. Remind that $p=0$ corresponds to the single configuration of
crossings representing the trivial knot.

\subsection{Unconditional distribution}
\label{sect:4.1}

\subsubsection{The mean complexity of the knots}
\label{sect:4.1.1}

The power of the polynomial invariant characterizes the complexity of a knot.
Hence, according to (\ref{eq:exp}) and definition of Kauffman invariant as
Potts partition function, the power $n$ has a sense of a free energy. Thus, for
long strips ($L_{h}\gg L_{v}$) neglecting the boundary effects produced by
shorter side $L_{v}$ we conclude that $n$ is proportional to $L_{h}$. This fact
directly follows from the additivity of the free energy. At the same time for
short strips $L_{h} \sim L_{v}$ the boundary effects cause the deviation from
linear behavior of $n$ upon $L_{h}$.

\begin{figure}[ht]
\begin{center}
\epsfig{file=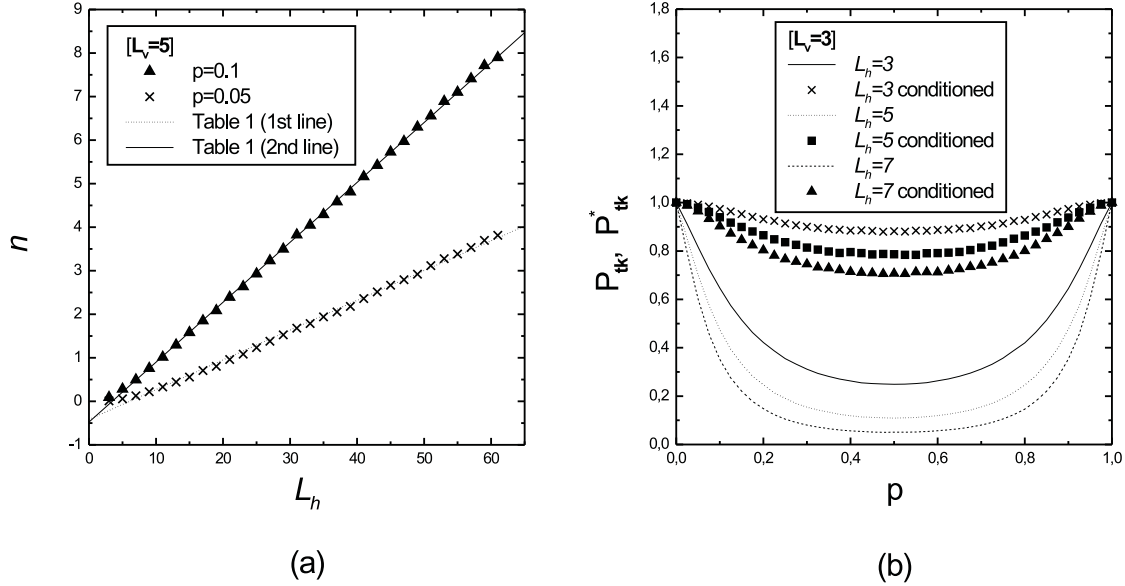,width=15cm}
\end{center}
\caption{a) Mean largest power $n(L_{h};p)$ of polynomial invariant
of a knot on a strip of width $L_{v}=5$ and length $L_{h}$ for
concentrations of impurities $p=0.05,0.1$; b) Probability of a trivial
knot $P_{tk}(L_{h};p)$ on the lattice $3 \times L_{h}$ for $L_{h}=3,5,7$
(lines) and conditional probability of trivial daughter (quasi)knot
$P^{*}_{tk}(L_{h};p)$ on a strip of width $L_{v}=3$ and length $L_{h}$
as function of $p$ under condition that the parent knot $L_{t}=2L_{h}+1$ is
trivial (crosses, squares, triangles).}
\label{fig:2}
\end{figure}

Our numerical computations remarkable confirm the above statement. The mean
complexity $n(L_h;p)$ of knots on the strip of width $L_{v}=3$ and results of
the approximation (Table~1) are
shown in
Fig.\ref{fig:2}a.

The corresponding data for the strip of the widths $L_{v}=3$ and  $L_v=5$
are shown in Table 1. We have averaged the results
over $30\,000$ realizations of topological disorder for each $L_{h}$. To
estimate the numerical error we have splitted the ensemble of realizations
onto 10 subsets of 3000 configurations each.

\begin{center}
{\bf Table 1}
\medskip

\begin{tabular}{|c|c|} \hline \multicolumn{2}{|c|}{$L_v=3$}
\\ \hline $p$ & $n(L_h;p)$ \\ \hline \hline $0.05$ & $0.0696(4)\times
L_h-0.4828(13)$ \\ \hline $0.1$ & $0.1381(4)\times L_h-0.51(1)$ \\ \hline
\end{tabular} \hspace{1mm}
\begin{tabular}{|c|c|} \hline
\multicolumn{2}{|c|}{$L_v=5$} \\ \hline $p$ & $n(L_h;p)$ \\ \hline \hline
$0.05$ & $0.1591(4)\times L_h-1.053(14)$ \\ \hline $0.1$ &
$0.3098(4)\times L_h-1.094(16)$ \\ \hline \end{tabular}
\end{center}

\subsubsection{The probability of a trivial knot formation}
\label{sect:4.1.2}

\begin{figure}[ht]
\begin{center}
\epsfig{file=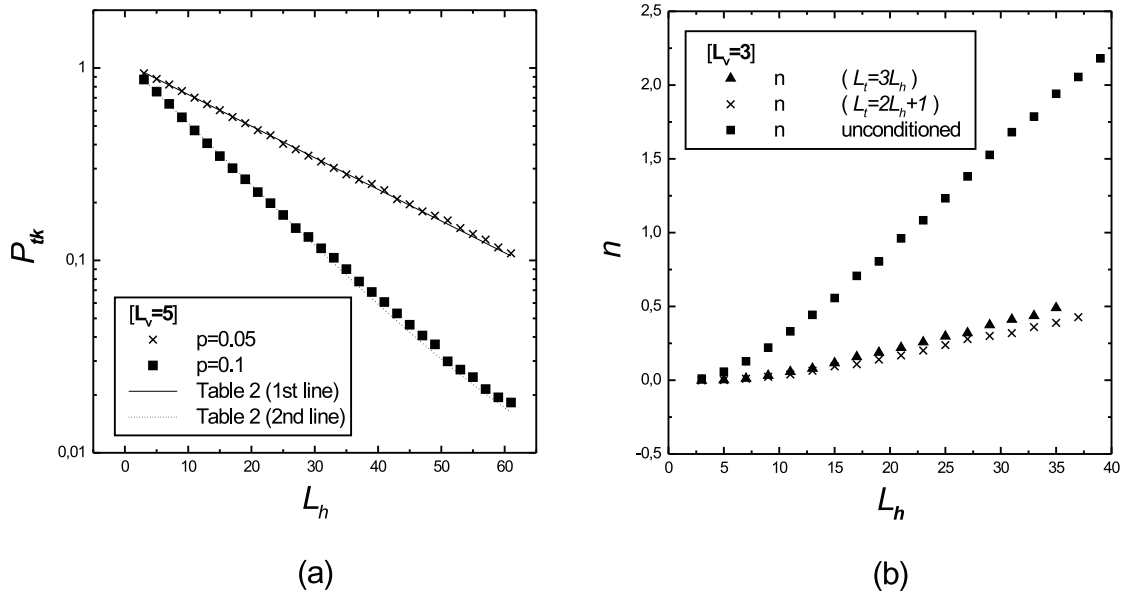,width=15cm}
\end{center}
\caption{a) Probability of a trivial knot $P_{tk}(L_{h};p)$ on a strip of
width $L_{v}=3$ and length $L_{h}$ for concentration of impurities
$p=0.05,0.1$; b) Complexity $n(L_{h};p)$ of a daughter (quasi)knots on
a strip of width $L_{v}=3$ and length $L_{h}$ for concentration of
impurities $p=0.05$ under condition that the parent knot $L_{t}=3L_{h}$
(triangles) and $L_{t}=2L_{h}+1$ (crosses) is trivial. For comparison, the
complexity of a random U--knots $n(L_{h};p)$ on a strip $3 \times
L_{h}$ is plotted by squares.}
\label{fig:3}
\end{figure}

The unconditional probability $P_{tr}(p;L_{h})$ of a trivial knot formation is
plotted in Fig.\ref{fig:2}b as a function of concentration of impurities $p$ on
strips $3\times L_{h}$ for $L_{h}=3,5,7$. The averaging is performed over
$10\times 10000$ samples. As it can be seen from Fig.\ref{fig:2}b the
probability of a trivial knot formation does not depend upon $p$, but only upon
$L_{h}$ for large $L_{v} \ge 9$ and intermediate values of $p \sim 0.5$. The
dependencies of the probability $P_{tk}(p;L_{h})$ upon $L_{h}$ on the strip of
fixed width $L_{v}=3$ for impurity concentration $p=0.05,0.1$ are shown in
Fig.\ref{fig:3}a in semi--logarithmic coordinates.

In the last case the averaging is performed over $10\times 3000$ samples. One
sees from Fig.\ref{fig:3}a that probability of a trivial knot formation
exponentially decays with the length of the strip. The results of corresponding
approximations are plotted in Fig.\ref{fig:3}a by lines. The same results of
approximation for $L_{v}=5$ are enclosed in the Table 2.

\begin{center}
{\bf Table 2}
\medskip

\begin{tabular}{|c|c|}
\hline \multicolumn{2}{|c|}{$L_v=3$} \\ \hline
$p$ & $P_{tr}(L_h;p)$ \\ \hline \hline
$0.05$ & $1.049(4)\exp(-0.03720(15)\times L_h)$ \\ \hline
$0.1$ & $0.957(22)\exp(-0.0673(5)\times L_h)$ \\ \hline
\end{tabular} \hspace{1mm}
\begin{tabular}{|c|c|}
\hline \multicolumn{2}{|c|}{$L_v=5$} \\ \hline
$p$ & $P_{tr}(L_h;p)$ \\ \hline \hline
$0.05$ & $1.155(7)\exp(-0.0794(3)\times L_h)$ \\ \hline
$0.1$ & $1.2305(22)\exp(-0.1568(11)\times L_h)$ \\ \hline
\end{tabular}
\label{tabcontr}
\end{center}

It follows from the Table 2 that the probability of a trivial knot
formation (i.e. of knots with $n=0$) is exponentially small for
sufficiently long strips.

\subsection{Conditional distribution (``Brownian Bridge'')}
\label{sect:4.2}

Let us investigate now the topological correlations of parts of a trivial knot.
Namely, consider a trivial ("parent") knot, generated on a strip of width
$L_{v}$ and length $L_{t}$. Cut a part of the strip of length $L_{h}$
($0<L_{h}<L_{t}$) and close the remaining open tails. In such a way we get the
new ("daughter") knot of the same width $L_{v}$ but of shorter length $L_{h}$.
In Fig.\ref{fig:1a}a,b the parent and a daughter knots are shown. All
lattice dimensions $L_v,\;L_t,\;L_h$ are taken to be odd.

We pay the main attention to the following problem. How the fact that the
parent knot $L_{v}\times L_{t}$ is trivial is reflected in the topological
properties of the daughter (quasi)knot $L_v\times L_h$? We consider two
cases
$L_h=L_t/3$ and $L_h=(L_t-1)/2$. To be more precise, we define the
conditional
probability that daughter (quasi)knot is trivial and find its mean complexity
under the condition that the parent knot is trivial.

The problem under consideration is typical for the theory of random walks.
Namely, in the theory of Markovian chains the conditional probability i.e. the
so called "Brownian Bridge" (BB) has repeatedly studied. The investigation of
statistics of BB supposes in first turn the determination of the probability
$P({\bf x},t|{\bf 0},T)$ that a random walk begins at the point ${\bf x}={\bf
0}$, visits the point ${\bf x}$ at some intermediate moment $0<t<T$ and returns
to the initial point ${\bf x}={\bf 0}$ at the moment $T$. The same question has
been investigated for BB on the graphs of free groups, on the Riemann surfaces
\cite{nesin1,bougerol} and for the products of random matrices of groups
$PSL(2,\Z)$ \cite{nesin2} and $PSL(n,\Z)$ \cite{letch}.

Its easy to understand, that topological problem under consideration can be
naturally interpreted in terms of BB. As it has been mentioned above (for
details see \cite{grne_alg,vasne}) the power $n$ of Jones--Kauffman polynomial
invariants defines the scale in the space of topological states of knots. That
allow us to compare knots and to talk about their respective
"complexity" or
"simplicity". Take a phase space $\Omega$ of all topological states of knots.
Select from $\Omega$ the subset $\Omega_{n=0}$ of knots with $n=0$. Cut a part
(say, one half, or one third) of each knot in the subset $\Omega_{n=0}$, close
the open ends and investigate the topological properties of resulting knots.
Just such situation was qualitatively investigated in \cite{grne_alg}, where
the authors have formulated the "crumpled globule" (CG) concept on the basis of
qualitative conjectures. The CG--hypothesis states the following. If the whole
knot is trivial, then the topological state of each its finite part is almost
trivial. Later on we formulate this statement in more rigorous terms and
confirm its by results of our numerical computations.

\subsubsection{The mean complexity of a (quasi)knot}
\label{sect:4.2.1}

We consider the strips of width $L_v=3$. The concentration of "impurities" is
set to $p=0.05$. The averaging is performed over $3\times 10^{4}$ selected
trivial knots. The mean complexities $n(L_{h})$ of "daughter" knots on the
strips $L_v\times L_h$ for $L_{v}=3$ under the condition that corresponding
"parent" knots on the strips $L_t=3L_h$ (triangles) and $L_t=2L_h+1$ (crosses)
are trivial, are shown on Fig.\ref{fig:3}b. We have compared the complexity of
BB (daughter) knots (triangles, crosses) with the unconditional (U) complexity
of  random knots (squares). Its easy to see, that complexity of the BB--knots
is sufficiently smaller than the complexity of U--knots. The functional
dependence of $n$ upon $L_{h}$ for  BB--knots shall be discussed later.

\subsubsection{The probability of the trivial (quasi)knot}
\label{sect:4.2.2}

Here we investigate the probability that a daughter knot on the strip $L_v
\times L_h$ is trivial under condition that it is a part of a trivial (parent)
knot on the strip $L_{t}=2L_h+1$. Results for $L_{h}=3,5,7$ are shown in
Fig.\ref{fig:2}b by crosses, squares and triangles respectively. The
unconditional probabilities of formation of the trivial knot are plotted
for
$L_{h}=3,5,7$ by lines in Fig.\ref{fig:2}b. As expected, the probability of a
(quasi)knot to be trivial becomes essentially higher under imposing condition
for this (quasi)knot to be a part of the trivial knot.

\subsection{Lyapunov exponents and the knot complexity}
\label{sect:5.1}

We are interested in the functional dependence of the complexity $n$ of the BB
(daughter) knot upon the lattice length $L_{h}$ for fixed width $L_v$. The
direct investigation of this question meets some technical difficulties and
to avoid them we shall use some properties of product of random matrices.
According to \cite{vasne} for a knot on the strip of width $L_{v}=3$ we
introduce the matrices
\be \label{eqtmatr}
\begin{array}{cc} P= (1+x)x^{-\frac{1}{2}}
\left( \begin{array}{cc} -x^{\frac{1}{2}} & 0 \\ (1+x)^{-1}x^{\frac{1}{2}}
& x^{-\frac{1}{2}} \\ \end{array} \right) & R= x^{\frac{1}{2}} \left(
\begin{array}{cc} -x^{\frac{1}{2}} & -(x+1)x^{-\frac{1}{2}} \\ 0 &
x^{-\frac{1}{2}} \end{array} \right) \\ \rule{0mm}{11mm} P^{*}=
(1+x)x^{-\frac{1}{2}} \left(\begin{array}{cc} -x^{-\frac{1}{2}} & 0 \\
(1+x)^{-1}x^{\frac{1}{2}} & x^{\frac{1}{2}} \\ \end{array} \right) & R^{*}=
x^{\frac{1}{2}} \left( \begin{array}{cc} -x^{-\frac{1}{2}} &
-(x^{-1}+1)x^{-\frac{1}{2}} \\ 0 & x^{\frac{1}{2}}
\end{array} \right) \end{array}
\ee
corresponding to ferro-- and antiferromagnetic bonds as shown in
Fig.\ref{fig:1b}d.

The polynomial knot invariants can be expressed as the product of  these
matrices. For example, the resulting matrix $A$ of the knot, shown in
Fig.\ref{fig:1b}a corresponding to the bond arrangement in
Fig.\ref{fig:1b}b,c
is:
\be \label{eq:tm}
A=PR^{*}P P^{*}RP^{*} PR^{*}P P R^{*}P P^{*} R P^{*}
\ee
From \cite{vasne} it follows, that the Jones polynome for the knot on the strip
of width $L_{v}=3$ reads:
\be \label{eq:kauff}
f_{K}(x)=(1+x)^{-N_{p}-1}\left(-\sqrt{x}
\right)^{N_{p}+1+N_{b}} \left((x+x^{-1}+2)a_{12}+(x+x^{-1}+2)^{2}a_{22}
\right)
\ee
where $a_{12}$ and $a_{22}$ are the elements of the last column of the
product
$A$ of the transfer matrices.

Let us investigate the relationship between the knot complexity $n$ and the
logarithm of the highest eigenvalue $n_{\lambda}$ of the product of transfer
matrices. Note that when calculating the knot invariant, one has the trivial
prefactor $d=(1+x)^{-N_{p}-1} \left(-\sqrt{x} \right)^{N_{p}+ 1+N_{b}}$, where
$N_{p}=(L_{v}+1)(L_{h}+1)/2$ is a number of Potts spin components and
$N_{b}=\sum \limits_{i,j} b_{i,j}$ is a sum over all $b_{i,j}$. So it is
convenient to add $\log d$ to the largest power $n_{\lambda}$ and introduce the
modified value $\tilde n_{\lambda}= n_{\lambda}-(N_{p}+1-N_{b})/2$. For the
strip on the lattice of width $L_{v}=3$ and concentration of impurities
$p=0.05$ the data for $\tilde n_{\lambda}$ in the interval $L_{h} \in [20,60]$
is well approximated by the linear function
\be \label{eq:lin}
\tilde n_{\lambda}(L_{h})=-1.16(2)+0.0741(6) \times L_{h}
\ee
This results establishes the linear dependence of the knot complexity upon
$L_{h}$.

\begin{figure}[ht]
\begin{center}
\epsfig{file=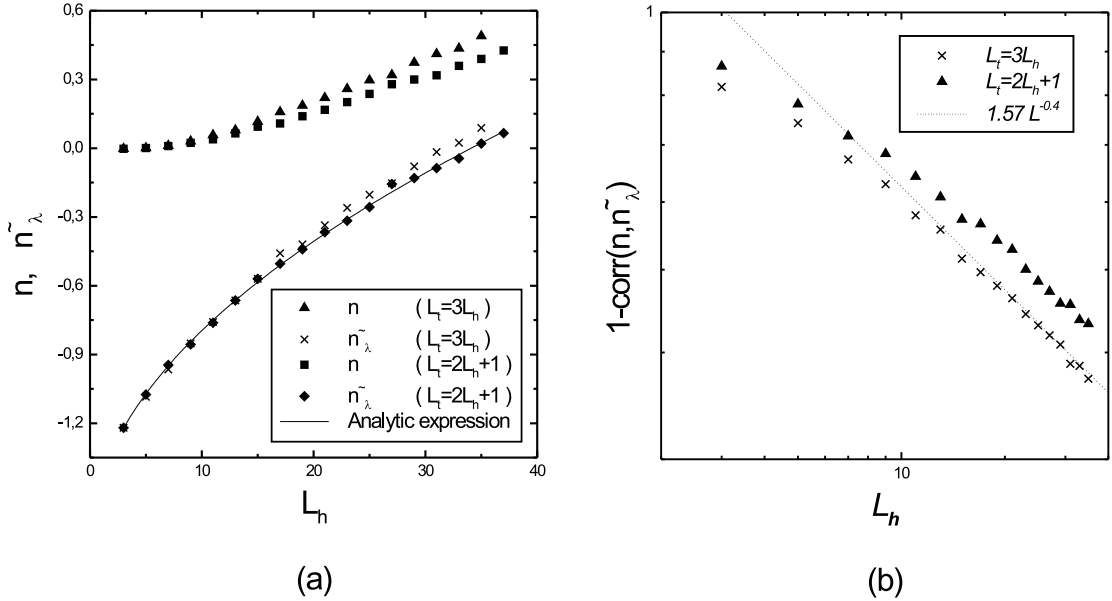,width=15cm}
\end{center}
\caption{a) Power of the Jones--Kauffman polynomial invariant
$n$ and the exponent of the maximal eigenvalue of the
product $\tilde n_{\lambda}$ of transfer matrices for BB--knots
$L_{t}=3L_{h}$ and $L_{t}=2L_{h}+1$; b) Correlation function ${\rm
corr}(n,\tilde n_{\lambda})$ as a function of $L_{h}$ for BB--knots
$p=0.05,0.1$, $L_{v}=3$.}
\label{fig:4}
\end{figure}

We have also computed $\tilde n_{\lambda}$ for BB--knot of length $L_{h}$ for
$L_{t}=3L_{h}$ and $L_{t}=2L_{h}+1$. The corresponding results are plotted in
Fig.\ref{fig:4}a.

These results are well approximated by the square-root function shown
in Fig.\ref{fig:4}a by a line:
\be \label{eq:sqr}
\tilde n_{\lambda}(L_{h})=-1.74(3)+0.300(22)\times L_{h}^{0.498(16)}
\ee
We have plotted in Fig.\ref{fig:4}a the data for $n$ for comparison as well.

The information about the behavior of the knot complexity we could obtain
analyzing the the correlation coefficient of the knot complexity $n$ and the
logarithm of the modified highest eigenvalue $\tilde n_{\lambda}$:
\be
{\rm corr}(n,\tilde n_{\lambda})=\frac{\left<n\,\tilde n_{\lambda}\right>-
\left<n\right>\left<\tilde n_{\lambda}\right>}{\sqrt{(\left<n^{2}\right>-
\left<n\right>^{2})(\left<\tilde n_{\lambda}^{2}\right>-\left<\tilde
n_{\lambda}\right>^{2})}}
\ee
The difference $1-{\rm corr}(n,\tilde n_{\lambda})$ is shown in
Fig.\ref{fig:4}b (the function $1.57 L^{-0.4}$ is added for comparison). We
see, that the correlation between $\tilde n_{\lambda}$ and $n$  approaches to
$1$ as $L_{h}$ tends to infinity. This behavior allows us to conclude that the
averaged knot complexity $n$ has the same dependence upon  $L_h$ as the mean
logarithm of the modified highest eigenvalue  $\tilde n_{\lambda}$.

\section{Conclusions}
\label{sect:5}

The correlation between $n$ and $\tilde n_{\lambda}$ established in the
previous section permits us to consider the distribution of Lyapunov exponents
for products of transfer matrices instead of distribution of largest power of
polynomial invariant.

Qualitatively the behavior (\ref{eq:sqr}) can be understood by considering
the limiting distribution of Lyapunov exponents of first $m$ matrices in a
product of $N$ random unimodular $2 \times 2$ matrices under the condition,
that the product as a whole equals the unit matrix. The explicit analysis
of the problem has been developed in \cite{nesin2}. In \cite{nesin2} the
authors studied the distribution of the Lyapunov exponent $\delta$ of first
$m=c N\; (0<c<1)$ matrices in the product of $N$ non-commuting
$PSL(2,\R)$--matrices whose elements are independently distributed in some
finite interval under the condition of the Brownian Bridge (i.e. under the
condition that the whole product equals the unit matrix). The exponent
$\delta$ obeys for $N\gg 1$ the following asymptotic behavior:
\be \label{eq:sqr2}
\delta \sim \sqrt{N}
\ee
(compare to (\ref{eq:sqr})). Without the Brownian Bridge condition (i.e. for
"open" chains of matrices) the standard F\"urstenberg result
\cite{fuerst_tut} has been reproduced
\be
\delta \sim N
\ee

In order to clarify the behavior (\ref{eq:sqr2}) we restrict ourselves
mainly with a qualitative consideration. The "Brownian Bridge" problem for
Markov chain of $N$ identically distributed noncommutative random matrices
may be reformulated in terms of a random walk in the space of the constant
negative curvature. Now it is well known \cite{nesin1,bougerol,letch} that
BB condition to return to the initial point "kills" the influence of the
curvature. The limiting distribution function turns to be the Gaussian
with zero mean. This fact can be illustrated by the following simple
computation.

Consider the random walk in a space of the constant negative curvature
$\gamma=-1$ (the Lobachevsky space) with the metric $ds^2=d\mu^2 +
\sinh^2\mu\,d\Phi^2$, where $d\Phi^2$ is a square of distance increment in the
space of angles. The probability of a path to start from the point $\mu=0$ and
to end at the time moment $t$ in a {\it particular} point located at
distance $\mu$ from the origin in the Lobachevsky space is well known:
\be \label{4pseud}
P(\mu,t) = \frac{e^{-t}}{8\sqrt{(\pi t)^3}}
\frac{\mu}{\sinh\mu} \exp\left(-\frac{\mu^2}{4t}\right)
\ee
(the diffusion coefficient $D$ is set to 1). For the first time
Eq.(\ref{4pseud}) was obtained in \cite{karp_gervas}. Respectively, the
probability to find a walker at time moment $t$ in {\it some} point at a
distance $\mu$ from the origin is:
\be \label{34lob}
{\cal P}(\mu,t)=P(\mu,t) {\cal N}_s(\mu)
\ee
where ${\cal N}_s(\mu)=\sinh^2\mu$ is an area of the sphere of radius $\mu$
in the Lobachevsky space.

The difference between $P$ and ${\cal P}$ is vanishing in the Euclidean space,
but becomes crucial in the non--Euclidean geometry. By using definition of the
Brownian Bridge we can easily calculate the conditional probability that
the
random walk starting and finishing at $\mu=0$ after $t$ first steps  visits
some point at distance $\mu $ from the origin in the Lobachevsky space.  This
probability for $N\to \infty$ reads:
\be \label{gauss}
{\cal P}(\mu,t|0,N)=\frac{P(\mu,t)P(\mu,N-t)\sinh^2\mu}{P(0,t)}=
\frac{N^{3/2}}{8\pi t^{3/2}(N-t)^{3/2}}\mu^2 \exp\left\{-\frac{\mu^2}
{4}\left(\frac{1}{t}+\frac{1}{N-t}\right)\right\}
\ee
Thus, we arrive at the Gaussian distribution with zero mean.

The behavior (\ref{gauss}) found for the random walks on the Riemann surface of
the constant negative curvature has the straightforward relation to the
conditional distribution of Lyapunov exponents of the products of
noncommutative random matrices. The Lobachevsky space $H$ may be identified
with the noncommutative group $SL(2,{\C})/SO(3)$. Let us take the
"Brownian Bridge" on ${\cal H}=SL(2,{\C})/SO(3)$. Namely consider the product
of $N$ random matrices $\widehat{\cal M}_k\in {\cal H}\; (0\le k\le N)$   under
the condition that the product $A=\prod_{k=1}^N \widehat{\cal M}_k$ is
identical to the unit matrix. We are interested in the distribution of
Lyapunov exponents $\delta$ for the first $m$ matrices in the product $A$.  The
stochastic recursion relation for the vector  $\disp {\bf W}_k$ reads:
\be \label{recurs}
{\bf W}_{k+1}=\widehat{\cal M}_k {\bf W}_k; \qquad {\bf W}_0={\rm const}
\ee
where ${\cal M}_k\in {\cal H}$ for all $k\in [0,N]$. The "Brownian Bridge"
condition means that
\be \label{bb}
{\bf W}_N={\bf W}_0
\ee
Consider for simplicity the case when each matrix in the product is close to
the unit one:
\be \label{norma}
\widehat{\cal M}_k=1+\widehat{M}_k;\qquad \mbox{norm}[\widehat{M}_k]\ll 1
\ee
The discrete dynamical equation (\ref{recurs}) under the condition
(\ref{norma}) may be replaced by the differential equation. Its stationary
distribution is defined by an appropriate Fokker--Plank equation for the
random walk in the Lobachevsky space. The distribution of the Lyapunov
exponent $\delta$ for the conditional product $A$ of random matrices is given
by (\ref{gauss}). So, under the conditions (\ref{bb}) and (\ref{norma}) for
$\delta$ we have recovered the usual Gaussian distribution. Hence the mean
value $\left<\delta_m\right>$ of the Lyapunov  exponent of any first $m=cN$
($c={\rm const}$, $0<c<1$) matrices has the square--root dependence
on $N$ ($N\gg 1$): $\left<\delta_m\right>\sim \sqrt{N}$.

This statement can be reformulated in topological terms. As it has been
shown, the correlation between the knot complexity and the Lyapunov
exponent tends to 1 for $L_{h}\to\infty$. Thus we can conclude that the
typical complexity $n$ of topological Jones--Kauffman invariant of a
daughter quasi--knot, obtained by cutting a part $N=L_{v}\times L_{h}$ of a
trivial parent knot of size $L_{v} \times L_{t}$, scales as
$$
n\sim \sqrt{N} \sim \sqrt{L_{h}}
$$
for $N \gg 1$. Contrary, for a random knot of size $N=L_{v} \times L_{h}$
without the Brownian Bridge condition the complexity scales as $n\sim N \sim
L_{h}$ (see \cite{vasne}) in agreement with the F\"urstenberg theorem
\cite{fuerst_tut}).

Therefore the relative complexity $\frac{n}{N}$ of a daughter quasi--knot,
which is a part of a trivial knot, tends to 0:
$$
\lim_{N\to\infty}\frac{n}{N}=\lim_{N\to\infty} \frac{{\rm const}\,
\sqrt{N}}{N}=0
$$

The Figs.\ref{fig:2}b and \ref{fig:3}b confirm our statement. Actually, the
mean complexity of a BB--knot is smaller than the typical complexity of a
random knot of the same size without the BB--condition. The probability for
a part of a trivial knot to be also trivial is sufficiently higher than the
corresponding unconditional probability for a random knot (of the same size) to
be trivial---see Fig.\ref{fig:2}b. Thus we directly validate the hypothesis
that the topological state of any daughter quasi--knot, which is a part of a
trivial parent knot, is almost trivial. Hence, the parts of a polymer chain in
the crumpled globule are practically unknotted in a broad range of scales.

The investigation carried out above could be considered as an indirect
verification of a hypothesis expressed in \cite{nechaev} concerning the
possibility to formulate some topological problems of collapsed polymer chains
in terms of path integrals over trajectories with prescribed fractal dimension
and without any topological ingredients. Namely, in ensemble of collapsed
polymer chains the essential fraction of trajectories has the fractal
dimension of dense packing state $D_{f}=d$ ($d$ is the space dimensionality).
Conversely there is a reason to assume that almost all paths in ensemble of
trajectories with fractal dimension $D_{f}=d$ (where $d\le 3$) are
topologically isomorphic to knots, close to trivial.

The problem of calculation of distribution for closed polymer chain with
topological constraints can be expressed as an integral over the set
$\Omega$ of closed paths with fixed topological invariant:
\be \label{5.1}
Z = \int\limits_{\Omega} D\{r\} e^{-H} = \int \ldots \int
D\{r\} e^{-H} \Delta [n],
\ee
where $D\{r\}e^{-H}$ is the integration with the Wiener measure and
$\Delta[n]$ extracts paths with the value of the power $n$ of
the topological Jones--Kauffman invariant equal to zero.

If our assumption is right, the integration over $\Omega$ in (\ref{5.1})
can be replaced by the integration over all paths without any
topological restriction but with special "fractal" measure $D_f\{r\}e^{-H_f}$:
\be \label{5.2}
Z = \int \ldots \int D_f\{r\} e^{-H_f}
\ee
The usual Wiener measure $D\{r\}e^{-H}$ is concentrated on the
trajectories with the fractal dimension $D_f=2$. We suppose, that for
description of statistics of ring unknotted polymer chains the measure
$D_f\{r\}e^{-H_f}$ with fractal dimension $D_f=d$ ($d\le 3$) should be used.

\begin{center}
{\bf Acknowledgments}
\end{center}
The work is partially supported by the RFBR grant 00-15-99302. O.A.V thanks
the laboratory LPTMS (Universit\'e Paris Sud, Orsay) for hospitality. We
appreciate the useful comments of J.-L. Jacobsen concerning the
realization of numerical algorithm used in our work. The authors are
grateful to Supercomputer Center (RAN) for available computational
resources.


\begin{thebibliography}{99}

\bibitem{lif} I.M. Lifshitz, JETP, {\bf 55} 2408 (1968)
\bibitem{lgkh} I.M. Lifshits, A.Yu. Grosberg, A.R. Khokhlov, Rev. Mod.
Phys., {\bf 50} 683 (1978)
\bibitem{grosbk} A.Yu. Grosberg, A.R. Khokhlov, {\it
Statistical physics of macromolecules} (New York: AIP Press, 1994)
\bibitem{gns} A.Yu. Grosberg, S.K. Nechaev, E.I. Shakhnovich,
J.Phys.(Paris), {\bf 49} 2095 (1988)
\bibitem{man} B.B. Mandelbrot {\it The Fractal Geometry of Nature}, (San
Francisco: Freeeman, 1982)
\bibitem{chu} B. Chu, Q. Ying, A. Grosberg, Macromolecules, {\bf 28} 180
(1995)
\bibitem{mcknight} W.L. Nachlis, R.P. Kambour, W.J. McKnight, Polymer, {\bf 39}
3643 (1994)
\bibitem{shakh} J. Ma, J.E. Straub, E.I. Shakhnovich, J. Chem. Phys.,
{\bf 103} 2615 (1995)
\bibitem{jones} V.F.R. Jones, Bull. Am. Math. Soc., {\bf 12} 103 (1985)
\bibitem{kauffman} L.H. Kauffman, Topology, {\bf 26} 395 (1987)
\bibitem{wu} F.Y. Wu, J. Knot Theory Ramific., {\bf 1} 47 (1992)
\bibitem{grne_alg} A.Yu. Grosberg, S. Nechaev, J. Phys. (A): Math. Gen.,
{\bf 25} 4659 (1992); A.Yu. Grosberg, S. Nechaev, Europhys. Lett.,
{\bf 20} 603 (1992)
\bibitem{vasne} O.A. Vasilyev, S.K. Nechaev, JETP, {\bf 93} 1119 (2001)
\bibitem{ligr} I.M. Lifshitz, A.Yu. Grosberg, JETP, {\bf 65} 2399 (1973)
\bibitem{nesin1} L.B. Koralov, S.K. Nechaev, Ya.G. Sinai,
Prob. Theory Appl., {\bf 38} 331 (1993) (in Rusian)
\bibitem{bougerol} P. Bougerol, Probab. Th. Rel. Fields {\bf 78} 193 (1988)
\bibitem{nesin2} S. Nechaev, Ya.G. Sinai, Bol. Soc. Bras. Mat., {\bf 21}
121 (1991)
\bibitem{letch} A.V. Letchikov, Russ. Math. Surv., {\bf 51} 49 (1996)
\bibitem{fuerst_tut} H. F\"urstenberg, Trans. Amer. Math. Soc., {\bf 198}
377 (1963); V.N. Tatubalin, Prob. Theory Appl., {\bf 10} 15 (1965),
Prob. Theory Appl., {\bf 13} 65 (1968)
\bibitem{karp_gervas} M.E. Gerzenshtein, V.B. Vasilyev, Prob. Theory Appl.,
{\bf 4} 424 (1959); F.I. Karpelevich, V.N. Tatubalin, M.G. Shur, Prob.
Theory Appl., {\bf 4} 432 (1959)
\bibitem{nechaev} A.Yu. Grosberg, S.K. Nechaev, Polymer topology, Adv. Polym.
Sci., {\bf 106} 1, in {\it Polymer Characteristics}, (Springer: Berlin,
1993); S.K. Nechaev, {\it Statistics of Knots and Entangled Random walks},
(WSPC: Singapore, 1996)

\end{thebibliography}
\end{document}